\documentclass[preprint,aps,showpacs]{revtex4}
\usepackage{graphics}
\begin{document}
\title{Harmonic decomposition to describe the nonlinear evolution of
Stimulated Brillouin Scattering}
\author{S. H\"uller, A. Maximov\footnote{Permanent address : Laboratory for Laser Energetics,
University of Rochester, 250 East River Road, Rochester NY 14623, USA},
and D. Pesme
}

\affiliation{Centre de Physique Th\'e{}orique, Ecole Polytechnique, CNRS UMR 7644, 91128 Palaiseau Cedex, France}
\date{\today}

\begin{abstract}
An efficient method to describe the nonlinear evolution of
Stimulated Brillouin Scattering in long scale-length plasmas is
presented. The method is based on a decomposition of the
hydrodynamics variables in long- and short-wavelength components.
It makes it possible to describe the self-consistent coupling
between the plasma hydrodynamics, Stimulated Brillouin Scattering,
and the generation of harmonics of the excited ion acoustic wave
(IAW). This description is benchmarked numerically and proves to
be reliable even in the case of an undamped ion acoustic wave. The
momentum transferred from the electromagnetic waves to the plasma
ions is found to induce a plasma flow which modifies the resonant
three wave coupling between the IAW and the light waves. A novel
picture of SBS arises, in which both IAW harmonics and flow
modification reduce the coherence of SBS by inducing local defects
in the density and velocity profiles. The spatial domains of
Stimulated Brillouin activity are separated by these defects and
are consequently uncorrelated, resulting in a broad and structured
spectrum of the scattered light and in a temporally chaotic
reflectivity.
\end{abstract}

\maketitle

The description of parametric instabilities in laser-produced
plasmas using simple coupled mode equations for three wave
interaction is no longer sufficient whenever the longitudinal
plasma waves are driven to large amplitudes. Then the
nonlinearities of the longitudinal wave can induce detuning with
respect to the three wave resonance. This is one of the reasons
usually invoked to explain why these simplified models
overestimate the scattering levels of Stimulated Brillouin
Scattering (SBS). In this article we concentrate on SBS, which is
the process by which the incident laser wave couples to an ion
acoustic wave (IAW) to give rise to a scattered transverse wave.
The generation of the harmonics due to the IAW fluid-type
nonlinearity \cite{1,candy,casa,rozmus,pfb3_3317} is already known to be able to
reduce significantly the SBS reflectivity when compared with the
results involving simply a linearized IAW. However, the previous
fluid-type models for SBS in Refs. \cite{1,candy,casa,rozmus}, aimed at
taking into account the IAW nonlinearity, were incomplete because they
did not properly describe the flow modification \cite{rose,ppcf44_B53}
caused by the incident transverse wave momentum deposition.
All the mentioned models \cite{1,candy,casa,rozmus,pfb3_3317} also
ignored multi-dimensional effects.
On the other hand, kinetic effects associated with particle
trapping \cite{Morales} give also rise to a nonlinear IAW frequency
shift and therefore modify the SBS nonlinear behavior.

In the present Letter, we reconsider the effect of the IAW
nonlinearities on SBS by accounting properly for the flow
modification caused by SBS. We first derive approximate equations
describing simultaneously the plasma hydrodynamics (i.e. the long
wavelength density and flow profiles), SBS, and the harmonic
generation of the excited IAW resulting from fluid-type
nonlinearity. Our method consists in decomposing the fluid
variables into long and short wavelength components, the latter
corresponding to the SBS generated IAW and its harmonics\cite{f3d}.
Our new code, based on this harmonic decomposition method, makes it
possible to describe plasmas of spatial sizes of the order of
realistic laser produced plasmas (of mm-size, typically), because
it does not resolve the IAW $\mu$m-scale.
We then continued a step further by checking the capacity of our
approach to account for kinetic effects effects by implementing in
the IAW propagator a nonlinear frequency shift modeling particle
trapping.\cite{Morales}

The transverse electric field is described by $E({\bf x},t) =
e^{-i \omega_0 t} \left( E_+ e^{i k_0 z} + E_- e^{-i k_0 z}\right)
+ c.c.$ where $E_+({\bf x},t)$ and $E_-({\bf x},t)$ are the
forward- and backward propagating light field components,
respectively, both enveloped in time and space with respect to the
light frequency $\omega_0$ and the wave number $k_0$. This wave
number is taken for a fixed reference plasma density $N_{eq}$
which yields, using the critical electron density $n_c$, $k_0^2 =
\omega_0^2 (1-N_{eq}/n_c)/c^2$. For the plasma density $n({\bf
x},t)$ and the velocity ${\bf v}({\bf x},t)$ we use a
decomposition separating the long-wavelength components $N_0({\bf
x},t)$ and ${\bf v}_0({\bf x},t)$ and the short-wavelength
components $n_p({\bf x},t)$ and ${\bf v}_p({\bf x},t)$, with
$|p|=1,2,...$,
\begin{eqnarray}
&& n = N_0 + \left( n_1 e^{i k_s z}+ n_2 e^{i k_s z} + .. + c.c. \right) , \ \ \nonumber \\
&& {\bf v} = {\bf v}_0 + \left({\bf v}_1 e^{i k_s z}+{\bf v}_2 e^{i k_s z} + .. + c.c.\right) , \ \nonumber
\end{eqnarray}
the first ($p$=0) representing the hydrodynamic evolution, and the terms with $p >$0, the fundamental
ion acoustic wave, $p=$1, excited by SBS, and its harmonics, $p>$1.
The reference wave number for the IAW is the wavenumber of backscattering, $k_s =2k_0$,
for which the ponderomotive force is proportional to
$\propto E_+ E_-^* \exp (i2k_0 z)$.

We use the paraxial approximation to reduce the wave equation for the total electromagnetic field
$E$ to two ``paraxial'' equations for $E_+({\bf x},t)$ and $E_-({\bf x},t)$,
\begin{eqnarray}
{\cal L}_{\rm par}(E_+) =-i(\omega_0/c N_{\rm eq})\left[n_1 E_- +\!(N_0-N_{\rm eq}) E_+\right] ,\label{par+}\\
{\cal L}_{\rm par}(E_-) =-i(\omega_0/c N_{\rm eq})\left[n_1^* E_+ +\!(N_0-N_{\rm eq}) E_-\right] ,\label{par-}
\end{eqnarray}
with the paraxial operator
${\cal L}_{\rm par}(E_{\pm}) = [\partial_t + c_{\pm}\partial_z
+\nu_t$ $-i (c^2/2\omega_0) \nabla^2_{\perp} ] E_{\pm}$, where
$c_+$ and $c_-$ stand for the group velocity of the
forward/backward propagating light, respectively, with $c_+ =c^2
k_0/\omega_0 =-c_-$, and $\nu_t$ denotes the damping of the
transversal waves. The right-hand-side (rhs) source terms in
equations (\ref{par+}) and (\ref{par-}) account for (i) resonant
3-wave coupling due to SBS, with the fundamental ion sound wave,
$n_1$, and for (ii) refraction on long-wavelength density
modifications, $N_0-N_{eq}$, causing e.g. self-focusing. In
comparison with the full wave equation without decomposition into
$E_{\pm}$, this model allows a considerably coarser spatial
resolution and thus much less numerical expense.

For the long-wavelength hydrodynamic component
we use the following set of equations,
assuming isothermal conditions, and
written in the conservative form on the left-hand side (lhs):
\begin{eqnarray}
&&\partial_t N_0 +\nabla N_0 {\bf v}_0 =\left(\partial_t n\right)_{\rm IAW}, \label{conti} \\
&&\partial_t \left( N_0 \bf{v}_0\right) +\nabla\left( N_0{\bf v}_0 {\bf v}_0\right)
+ c_s^2\nabla N_0 = \  \label{moment} \\
&& \hspace{1.5cm} -N_0 c_s^2 \nabla U_0 +\left( \partial_t n {\bf v}\right)_{\rm IAW}, \nonumber
\end{eqnarray}
where the rhs source terms, $(\partial_t n)_{\rm IAW}$ and $(
\partial_t n {\bf v})_{\rm IAW}$, describe the momentum transfer
into the flow due to the IAW excitation by SBS, with $\left(
\partial_t n {\bf v}\right)_{\rm IAW} \equiv 2 c_s \left( 2\nu_{s1} -
{\bf v}_0\cdot\nabla\right) |n_1|^2/N_0$, and $\left(\partial_t
n\right)_{\rm IAW}\equiv -2 c_s \nabla\left( |n_1|^2 /N_0\right)$
. The ponderomotive force is given by $\nabla U_0 =\epsilon_0
\nabla (|E_+|^2 + |E_-|^2 )/n_c T_e$. The equations describing the
IAW driven by SBS, $n_1$, and its harmonics, $n_{l> 1}$ (using the
convention $n_{-l}=n_l^*$ for the complex conjugate) can be
written, in the so-called weak coupling regime, as follows
\begin{eqnarray}
&&\left[\partial_t + \nu_{sl} + i\omega_l +(v_{0z} +v_{gl})\partial_z
-i (c_s/2l k_s) \nabla^2_{\perp}\right]n_l =
\nonumber \label{iaw}  \\
  && \hspace{.3cm} -i (k_s c_s/2) N_0 \left[
\delta_{l,1} \left(\epsilon_0 E_+ E_-^* / n_c T\right) +  2 Q_l /N_0^2 \right]
\end{eqnarray}
with $Q_l\!=\!(l/2)\sum n_h n_{l-h}$ for $h\!\neq\!0$ and
$l\!\neq\!h$, where $c_s =[(Z T_e+3T_i)/M_i]^{1/2}$ is the IAW
speed (with $Z$ and $M_i$ as the ion charge and mass), $v_{0z}$
the $z$-component of the flow ${\bf v}_0$; $v_{gl}$ and $\omega_l$
denote the group velocity and the ``local'' frequency of the
$l$-th IAW harmonic, both accounting for the dispersion due to
Debye shielding increasing with the harmonic order. They are given
by $v_{gl} = c_s (1+l^2 k_s^2 \lambda_D^2)^{-3/2}$ and by
$\omega_l(z) = \omega_s(l k_s c_s) + l k_s v_{0z}(z)$ with the IAW
frequency $\omega_s(k) = k c_s (1+ k^2 \lambda_D^2)^{-1/2}$.
Equations (\ref{par+})-(\ref{iaw}) describe what we call the
harmonic decomposition model. They form a closed system describing
SBS in a temporally and spatially evolving plasma. They can be
shown to conserve momentum \cite{ppcf44_B53} at the lowest order in
$1/(k_s \ell_{\parallel})$ and in $1/(k_s \ell_{\perp})^2$ (with the
inhomogeneity length $\ell_{\parallel}=|\partial_z v_0/v_0|^{-1}$
and $\ell_{\perp}=|\nabla_{\perp} v_0/v_0|^{-1}$).

In the following we emphasize the particular importance (i) of the
SBS-induced flow modification, originating from the rhs term of
Eq. (\ref{conti}) as well as of the term $(\partial_t n {\bf
v})_{\rm IAW}$ on the rhs of Eq. (\ref{moment}), and (ii) of the
IAW harmonic generation described by the coupling terms
$\propto\sum n_h n_{m-h}$ in the rhs of Eq. (\ref{iaw}). In order
to stress the effect of each mechanism, we neglect for simplicity
the IAW damping, (while being aware that the IAW damping
coefficient is usually of the order of a few percent of the IAW
frequency). Indeed, the SBS-induced flow modification due to
momentum transfer, first pointed out by Rose in Ref.\cite{rose},
cannot be ignored in the regime of absolute instability
corresponding to weak IAW damping, because it is just in this regime
that the stationary 1D limit of Eqs. (\ref{conti}) and (\ref{moment})
exhibits the most pronounced flow modification. Namely, the
generation of the backscattered light gives rise to a transfer of
momentum to the bulk plasma in the spatial domain of SBS activity.
This momentum transfer results in a decrease $\Delta v \equiv
v_{0,out} - v_{0,in}<0$ of the flow $ v_0$ in the direction of
propagation of the laser, the net flow decrease being given by
$\Delta v \simeq -2R_{\rm SBS}(2\epsilon_0 |E_+|^2/N_{eq}
T_e)(1-N_{eq}/2n_c)$. Here, $R_{\rm SBS}$ denotes the SBS
reflectivity corresponding to the considered SBS active region.

We have performed simulations on the basis of equations
(\ref{par+})-(\ref{iaw}) and expanded the IAW up to its 3rd
harmonic, resulting in a set of equations for $n_1$, $n_2$, and
$n_3$, with the rhs terms $Q_1 = n_2 n_1^* + n_3 n_2*$, $Q_2 =
n_1^2 + 2 n_3 n_1^*$, and $Q_3 = 3 n_2 n_1$. We did not observe
any significant changes when harmonics above the 3rd order were
retained, while restricting to less than 3 harmonics led to
important differences.
\begin{figure}
\resizebox{8.5cm}{!}{\includegraphics{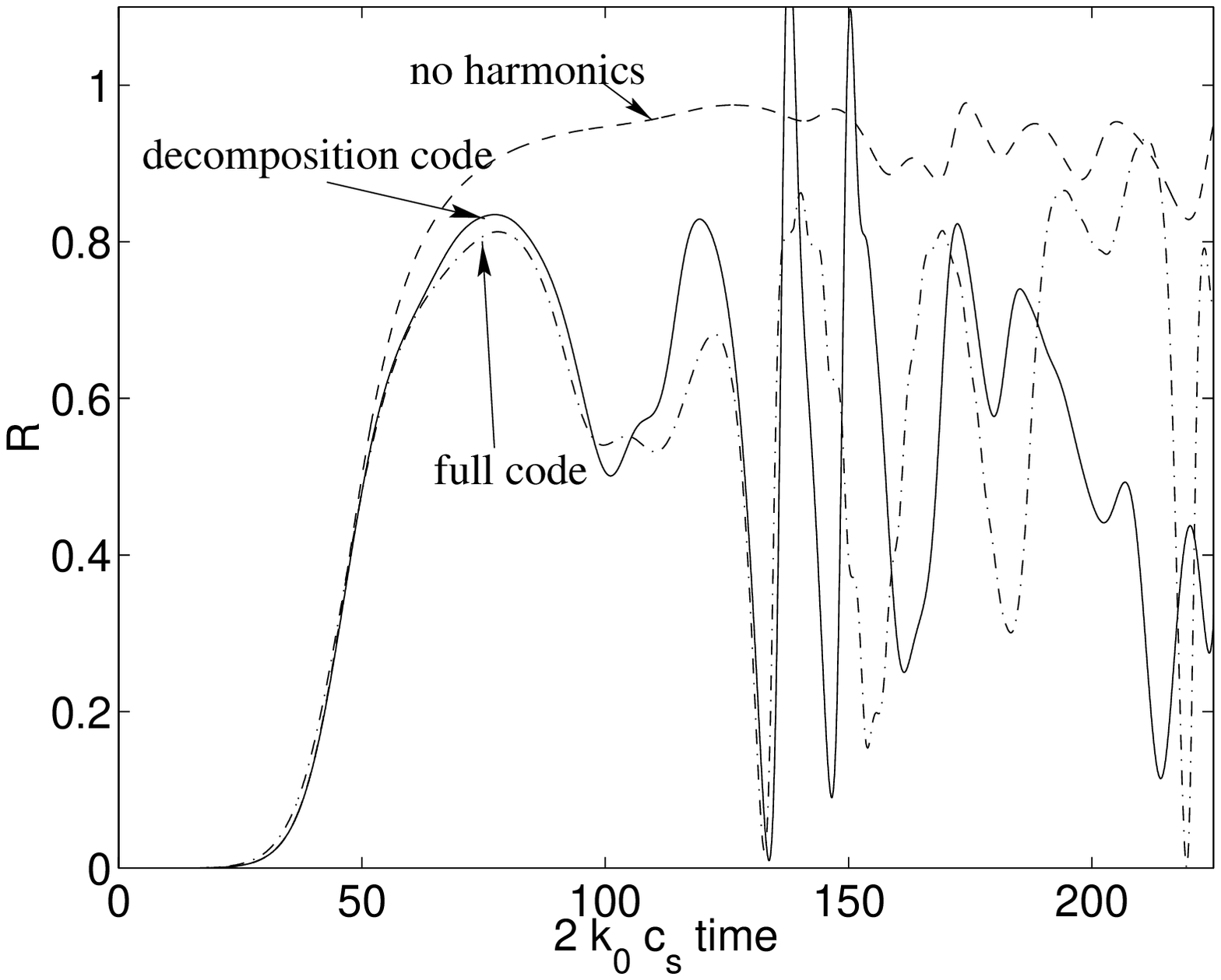}}
\caption{SBS reflectivity $R_{\rm SBS}$ versus time for the case of
an undamped IAW with the parameters $I_L=2.5\cdot 10^{14}$W/cm$^2$ for
$\lambda_0=1.064\mu$m at $T_e=1$keV,
$N_0/n_c =$0.1 (taken at center), $2 k_0\lambda_D =$0.27,
$L_{\rm ini}\simeq 160\lambda_0$.
The solid line is obtained from the decomposition code considering
all terms, the dashed line from the decomposition code disregarding
higher IAW harmonics, and the dash-dotted line from the ''full'' code.}
\end{figure}
At this stage of our study we restricted ourselves to
one-dimensional (1D) simulations in order to benchmark our
harmonic decomposition code against a ``complete'' 1D code which
does not make the decomposition corresponding to Eqs. (1)-(3).
This latter code solves Helmholtz's equation for the total
electric field $E(z,t)$ on the first hand, and the system of fluid
equations for continuity and momentum, with the complete
ponderomotive force, $\nabla |E(z,t)|^2$, as a source term, on the
second hand.
Here, in 1D, the operator $\nabla$ reduces to the partial derivative
${\bf e}_z \partial_z$.

To ensure equivalent boundary and initial conditions we have
considered a realistic case similar to an ``exploding foil'',
where an initially heated plasma expands starting from an almost
box-like density profile, with smooth shoulders, in the interval
$z_1 < z < z_2$ along the laser axis. The plasma profile, with the
initial plateau width $L_{\rm ini}\simeq 160\lambda_0$,
successively undergoes rarefaction from each side, so that the
velocity profile eventually tends to a monotonous curve varying
from negative to positive values with $v_0 =0$ in the center. The
simulation box is chosen in such a way that the rarefaction of the
profile does not significantly change the boundary conditions for
the light fields at the entrance ($z_{\rm ent}=0 < z_1$) and the
rear side ($z_{\rm rear}>z_2$). The total box size is $z_{\rm
rear}=2000/k_0\simeq 320\lambda_0$, where $\lambda_0=2\pi/k_0$
denotes the laser wavelength. The boundary condition for the
incident light at $z=0$ is a constant, $E_+(0)=const$, whereas the
backscattered light is seeded with a noise source at the level
$\langle |E_-(z\!=\!z_{\rm rear})|^2\rangle \sim 10^{-6}
|E_+(z\!=\!0)|^2$ and with a spectral bandwidth sufficiently
larger than the IAW frequency, in order to cover all possible SBS
resonances in the profile. In the density profile wings left and
right of the central plateau (for times $t < L_{\rm ini}/2 c_s$),
the plasma is strongly inhomogeneous in velocity and density so
that SBS is inhibited by the strong flow gradient.

We carried out our simulations in the absolute instability regime
of SBS with undamped IAWs, both to examine the role of flow due to
momentum transfer, and to benchmark the robustness of our
decomposition code. Notice that in the case of completely undamped
IAWs the SBS saturation level is, according to Refs.
\cite{maximov,Fuchs}, independent of the noise level. For the chosen
electron density $N_{eq}/n_c =0.1$ and for the plasma length
indicated above, the standard three-wave interaction model for
undamped IAWs \cite{maximov,Fuchs} predicts a steep increase in the SBS
reflectivities $R_{\rm SBS}$ as a function of the laser intensity,
varying from $R_{\rm SBS}<<1$ for small laser intensities to
$R_{\rm SBS}\simeq 1$ for normalized laser intensities above
$a_0^2 = \epsilon_0 |E_0|^2/n_c T_e \simeq 0.003$, with $E_0$
denoting $E_0 =E_+(z=0)$.
\begin{figure}
\resizebox{8.5cm}{!}{\includegraphics{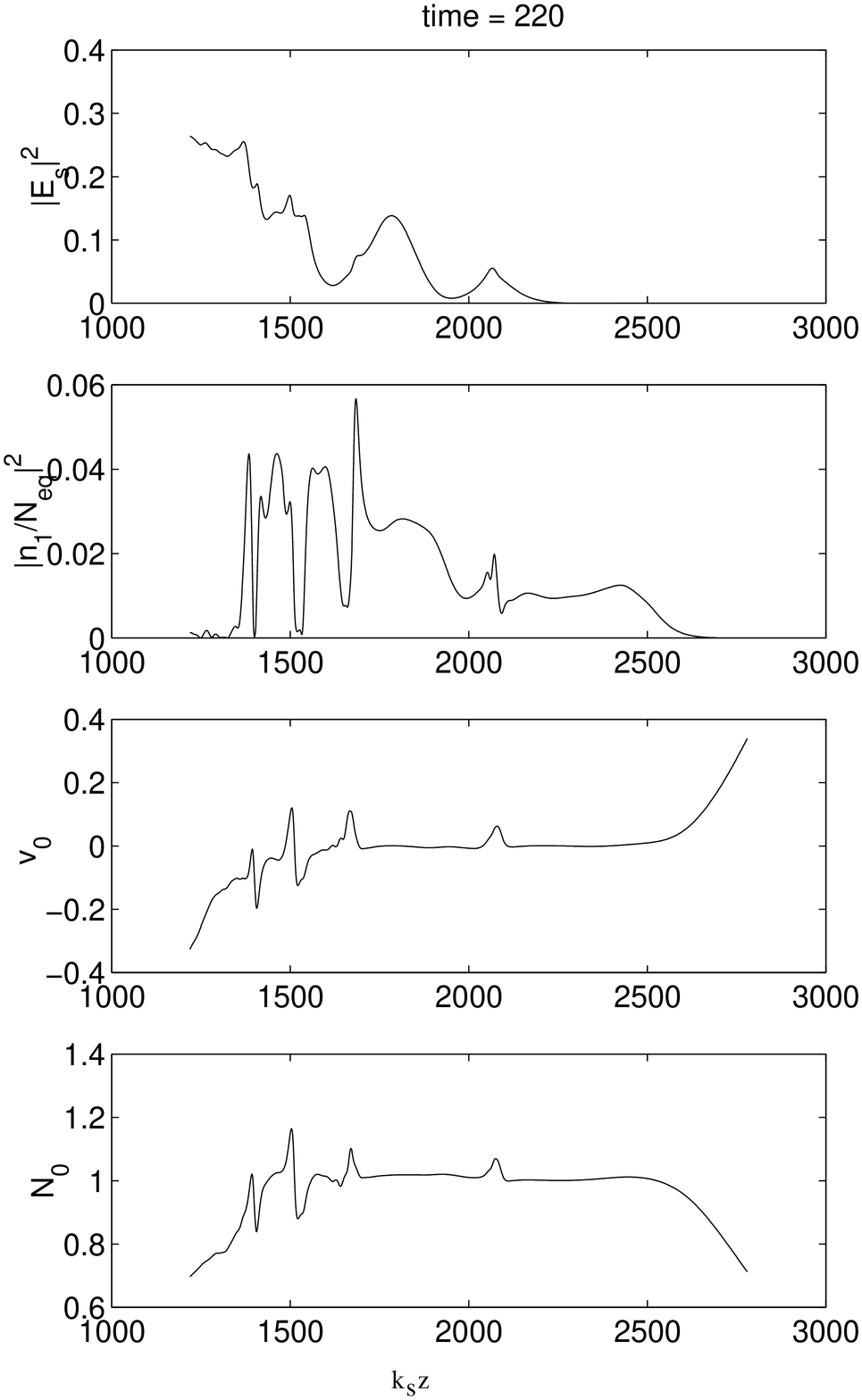}}
\caption{Spatial profiles, taken at $2 k_0 c_s t=220$,
of the backscattered intensity (upper subplot),
the fundamental IAW square amplitude $|n_1|^2$, the flow
velocity $v_0$, and the density $N_0$ inside the shoulders
of the exploding foil profile.}
\end{figure}
Our simulations comparing the decomposition code and the
``complete'' code show very good agreement, even for the extreme
case shown in Fig. 1, corresponding to the plasma parameters
mentioned above and to $a_0^2 = 0.025$ and $2 k_0\lambda_D =0.27$
(corresponding to an electron temperature $T_e = 1$keV and a laser
intensity $I_L =2.5\cdot 10^{14}$W/cm$^2$ at
$\lambda_0=1.064\mu$m), for which the reflectivity would be 99\%
in the absence of any IAW nonlinearity or flow modifications. For
lower intensity values, the agreement is even more striking. This
excellent agreement between the two codes gives us confidence in
the robustness of the harmonic decomposition description. In the
simulation presented in Figs. 1 and 2, the maximum amplitudes of
the harmonics remained below the validity condition for harmonic
expansion, namely $|n_l/N_{eq}| < 6^{1/2} (l k_s\lambda_D)^2$ for
$l=1,2,3$.

It can be observed, in the spatial profiles shown in Fig. 2 for
the backscattered intensity $|E_-|^2$, the fundamental IAW
amplitude $|n_1|$, the flow $v_0$, and the plasma profile $N_0$,
that the IAW behavior and flow modifications are entirely
connected with the existence of ``defects'' in these spatial
profiles and with a non-monotonous character in space (see also
Ref.\cite{rozmus}). Namely, SBS develops in distinct spatial domains,
interrupted by phase defects, which originate in the density
profile shoulders corresponding to the low density plasma on the
laser entrance side, and which then propagate into the profile
plateau.
Thus the SBS activity in each spatial domain appears to be
uncorrelated, due to their different origin in the inhomogeneous
velocity profile $v_0$. This feature reflects in the structured
nature of the backscattered light temporal spectrum, shown in Fig.
3 in which distinct peaks appear, and, consequently, in the
temporally chaotic behavior of the reflectivity.

Our decomposition description makes it possible to discriminate
the relative importance of the various effects contributing to the
nonstationary behavior in the SBS reflectivity, as seen in Fig. 1.
By suppressing parts of these effects in different runs, we have
found that the most important effect is the excitation of the IAW
harmonics: namely, retaining the harmonic excitation and
neglecting the SBS-induced flow modification lead to results that
remain in reasonably good agreement with the exact model, whereas,
retaining the flow modification, but ignoring the harmonics leads
to unphysically high levels of IAW amplitudes. It follows from
these observations that a realistic modeling of SBS requires the
proper description of the IAW harmonics. The 1D simulations
presented here would correspond to the SBS development in a long
laser hot spot. We have recently carried out 2D simulations which
confirm the relevance in 2D of the scenario described above
whenever the hot spot focus is not far (less than approximately one
Rayleigh length) from the transition between the inhomogeneous and
the homogeneous domain of the plasma density profile (i. e. the
shoulder of the expanding plasma in our case).

Increasing the laser intensity induces stronger IAW amplitudes at
which ion and/or electron kinetic effects take place.
We have included phenomenologically weak ion kinetic effects in
our decomposition model by adding a non-linear frequency shift of
the form $-i\eta |n_1/N_{eq}|^{1/2}$, \cite{Morales} in the
propagator appearing in the lhs of Eq. (\ref{iaw}) describing the
evolution of the IAW fundamental component and of its harmonics.
Although this is subject of work in progress, let us mention that
we have solved numerically the corresponding equation Eqs.
(\ref{par+})-(\ref{iaw}), and we find that for a positive and
sufficiently large $\eta$ coefficient ($\simeq 0.5\ldots 0.7$),
this shift can smooth out the effect induced by the harmonics and
the flow,
in a way such that (i) the ``defects'' are less pronounced and
(ii) the SBS reflectivity diminishes, but without exhibiting a
strong nonstationary behavior.
\begin{figure}
\resizebox{8cm}{!}{\includegraphics{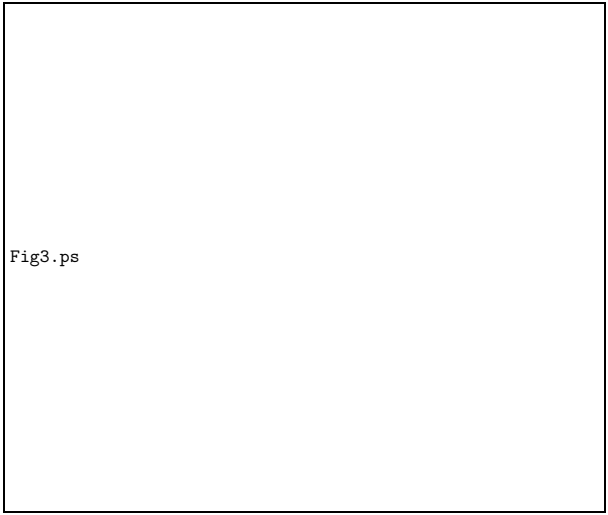}}
\caption{Spectrum of the backscattered light, corresponding to Fig. 1.
The frequency is shifted with respect to the incident laser frequency
$\omega_0$.}
\end{figure}
In conclusion, we have shown that the SBS modeling presented here,
based on a harmonic decomposition of the hydrodynamics variables,
represent a promising way to describe laser plasma interaction in
long scale-length plasmas. We have benchmarked our code based on
the harmonic decomposition in the extreme limit of the absolute
instability regime by neglecting the IAW damping. A novel picture
of SBS arises in which an incoherent superposition of scattered
light generated in distinct spatial domains in the velocity
profile leads to a nonstationary character of the SBS reflectivity
and to a significant reduction in the time averaged reflectivity.
This harmonic decomposition description appears to be sufficiently
robust and versatile to allow further sophistication by including
additional mechanisms such as kinetic effects via an
amplitude-dependent nonlinear frequency shift. We currently work
on a generalization of the harmonic decomposition method in order
to include the subharmonic IAW decay.

The numerical simulation were carried out thanks to the access to
the facilities of IDRIS at Orsay, France. The authors would like
to acknowledge fruitful discussion with  L. Divol, J. Myatt, 
C. Riconda, H. A. Rose, and W. Rozmus.

\end{document}